\documentclass[12pt]{iopart}
\usepackage{amssymb}
\usepackage{bm}
\usepackage[dvips]{graphicx}
\newcommand{\be}{\begin{equation}}
\newcommand{\en}{\end{equation}}

\newcommand{\Str}{\,{\mathrm{Str}}}
\newcommand{\diag}{\mathrm{diag}}
\newcommand{\re}{\mathrm{Re}}
\newcommand{\erfi}{\mathrm{erfi}}
\newcommand{\pQ}{Q^{\prime}}
\newcommand{\lb}{\lambda_B}
\newcommand{\lf}{\lambda_F}
\newcommand{\lbp}{\lambda_B^{\prime}}
\newcommand{\lfp}{\lambda_F^{\prime}}
\newcommand{\mb}{\mu_B}
\newcommand{\mf}{\mu_F}
\newcommand{\mbp}{\mu_B^{\prime}}
\newcommand{\mfp}{\mu_F^{\prime}}
\newcommand{\Pomega}{\Omega^{\prime}}

\newcommand{\etap}{\eta^{\prime}}
\newcommand{\hq}{\hat{Q}}
\newcommand{\td}{\tilde{D}}
\newcommand{\tl}{\tilde{\lambda}}
\newcommand{\tq}{\tilde{Q}}
\newcommand{\prl}{Phys. Rev. Lett.}

\newcommand{\prb}{Phys. Rev. B}

\begin{document}
\title{Virial expansion of the non-linear sigma model in the strong coupling limit}
\author{A. Ossipov}
\address{School of Mathematical Sciences, University of Nottingham, Nottingham NG7 2RD, United Kingdom} 

\begin{abstract}

We develop a perturbative approach to study the supersymmetric non-linear sigma model characterized by a generic coupling matrix in the strong coupling limit. The method allows us to calculate explicitly the moments of the eigenfunctions and the two-level correlation function in the lowest order of the perturbative expansion. We find that the obtained expressions are equivalent to the results derived before for the corresponding random matrix ensembles. Such an equivalence is elucidated and generalized to all orders of the perturbative expansion by mapping the sigma model onto the field theory describing the almost diagonal random matrices.

\end{abstract}

\maketitle

\section{Introduction}

One of the most powerful approaches to study complex quantum systems is based on the field-theoretical formulation. The non-linear $\sigma$-model, as an effective field theory describing long-distance physics, was first derived by Wegner \cite{W79, SW80} in the context of disordered quantum systems using the replica trick. An alternative formulation of the $\sigma$-model  given by Efetov \cite{E83, Efetov} employed supersymmetry method making use of both commuting and anticommuting variables. The latter version of the $\sigma$-model can be successfully applied not only for problems, which can be solved on the perturbative level, but allows also for a non-perturbative treatment of the theory. Moreover, the supersymmetric $\sigma$-model can be derived as an effective field theory for various random matrix ensembles. In particular, it was used to prove that the statistical properties of the energy levels in small metallic disordered grains can be described by the Wigner-Dyson random matrix theory (RMT) \cite{E83, Efetov}.

The partition function of the generic supersymmetric non-linear $\sigma$-model on a lattice 
\be\label{partition}
Z=\int DQ\; e^{-\frac{1}{t}\sum_{m\neq k}U_{mk}\Str Q_mQ_k-\sum_{n}S_0[Q_n]}
\en
is defined in terms of supermatrices $Q_m$ associated with $m$th lattice site, a symmetric coupling matrix $U_{mk}>0$ and a coupling constant $t$. $S_0$ comprises  all the local terms of the action including source terms. An explicit parametrization of $Q$ and of the measure $DQ=\prod_{m}dQ_m$ is given in \ref{app1}.

An important feature of the $\sigma$-model is the fact that the super-matrix field $Q$ satisfies a non-linear constraint $Q^2=1$. It is this constraint, which makes generally the analysis of the  $\sigma$-model to be highly non-trivial. One possible way to tackle the problem is to use the perturbation theory of the non-linear $\sigma$-model in the metallic regime. The latter is characterized by a small value of the coupling constant $t$, which corresponds to high electrical conductivity in the case of disordered conductors. In this regime the statistical properties of energy levels and eigenfunctions are close to the corresponding ones of the Wigner-Dyson RMT. A large number of important and interesting results have been obtained following this route (see \cite{M00} and references therein).

The opposite limit of the strong coupling has received much less attention. One of the reasons for that is the fact that in many cases the mapping between original microscopic models and the $\sigma$-model can be justified only at small values of the coupling constant. At the same time it is possible to construct microscopic models (such as for example $n$-orbital Wegner model \cite{W80}), which can be mapped onto the $\sigma$-model at all values of the coupling constant. This situation can be realized, for example, in a cluster of coupled disordered grains or chaotic systems, provided that each element of a cluster is well described by the Wigner-Dyson RMT. In such a case the properties of the $\sigma$-model in the weak and in the strong coupling limits are equally important. Moreover, some properties of the $\sigma$-model in the strong coupling limit might be of interest independently of whether or not it can be directly derived from a microscopic model. This is especially true for critical exponents of the $\sigma$-models which remain critical at large values of the coupling constant, as critical exponents are the same for all microscopic models belonging to the same universality class \cite{MFME06}.

Another reason, which makes the strong coupling limit to be less explored, is that technically this problem is more challenging, as the standard perturbation theory approach can not be applied there. This is in contrast to the weak coupling limit, where after separating out slow and fast modes, the latter can be integrated out order by order using the Wick's theorem. 

In this paper we develop a perturbation theory for the supersymmetric non-linear $\sigma$-model defined on a lattice in the the strong coupling limit. The general idea of our approach is similar to the one of the virial expansion method introduced in Ref.\cite{YO07} for studying almost-diagonal random matrices. After presenting a construction of the virial expansion for the  generic $\sigma$-model, characterized by an arbitrary coupling matrix $U$, we calculate explicitly the first non-trivial terms of the expansion for two different quantities: the moments of the eigenfunctions $I_q$ and the two-level correlation function $R_2(\omega)$. Then we compare the obtained results with the corresponding expressions derived for almost-diagonal random matrices. Our main conclusion is that the expressions for $I_q$ and $R_2(\omega)$ are in one-to-one correspondence with the random matrix results. This fact can be traced back to the properties of the target space of $Q$ and its counterpart discussed in Ref.\cite{OK06}. We show that such a correspondence can be established at any order of the virial expansion under very general assumptions.

The layout of this paper is as follows. In Section \ref{sec_virial}, we present the general framework for the virial expansion formulated for the generic non-linear $\sigma$-model. Section \ref{sec_moments} describes the calculations of the moments of the eigenfunctions $I_q$ and the distribution function of the eigenfunction intensities $P(z)$ in the lowest non-trivial order of the coupling constant. Similar calculations for the two-level correlation function  $R_2(\omega)$ are presented in Section \ref{sec_correl}. In Section \ref{sec_power} we apply the general formulas to the power-law non-linear $\sigma$-model and compare the obtained expression with the results known for the almost-diagonal random matrices. Finally, in Section \ref{conclusions} we discuss the origin of the results and demonstrate the equivalence between the $\sigma$-model description and the supersymmetric approach developed in Ref.\cite{YO07} in the strong coupling limit.

\section{Virial expansion method}\label{sec_virial}

The main idea of the virial expansion method is to represent the ``interaction'' part of the action $\frac{1}{t}\sum_{m\neq k}U_{mk}\Str Q_mQ_k$ as a sum over contributions characterized by a given number 
of ``interacting'' matrices $Q_m$ \cite{YO07}. To this end, we denote by $S[Q_m,Q_k]\equiv\frac{2}{t}U_{mk}\Str Q_mQ_k$ the sum of two terms which couple $Q_m$ with $Q_k$. Then the ``interaction'' part of the partition function can be expanded as follows
\begin{eqnarray}
e^{-\sum_{m>k}S[Q_m,Q_k]}=\prod_{m>k}e^{-S[Q_m,Q_k]}=\prod_{m>k}(1+F_{mk})=\nonumber\\
1+\sum_{m>k}F_{mk}+\sum_{m>k>p}(F_{mk}F_{kp}+F_{mp}F_{kp}+F_{mk}F_{mp}+F_{mk}F_{kp}F_{mp})+\dots,
\end{eqnarray}
where $F_{mk}$ stands for $e^{-S[Q_m,Q_k]}-1$. The first term of the above expansion is trivial. In the second one the contributions of pairs of supermatrices are collected together,  the third one takes into account the contributions of triplets and so on. As a result we obtain the virial expansion of the partition function:
\be\label{Z_expan}
Z=\int DQ\; e^{-\sum_{n}S_0[Q_n]}+\sum_{m>k}\int DQ\;\left(e^{-S[Q_m,Q_k]}-1\right) e^{-\sum_{n}S_0[Q_n]}+\dots.
\en
In the case of the disordered metallic grains the above expansion is useful, when the grains are weakly coupled.

This general scheme is elaborated in the next two sections, where it is used to calculate the moments of the eigenfunctions $I_q$ and the two-level correlation function $R_2(\omega)$ in the unitary symmetry class ($\beta=2$).

\section{Moments of the eigenfunctions}\label{sec_moments}

The expression for the moments $I_q$ of the eigenfunctions (also known as generalized inverse participation ratios) in terms of the $\sigma$-model reads \cite{FM_int,FM95}
\be
I_q=\lim_{\eta\to 0}\frac{\eta^{q-1}}{4N}\int DQ \sum_n \Str^q\left(Q_n k \Lambda\right)e^{-\frac{1}{t}\sum_{m\neq p}U_{mp}\Str Q_mQ_p-\eta\sum_{m}\Str\Lambda Q_m}, 
\en
where $k=\mathbf{1}\otimes\diag(1,-1)$ and $\Lambda=\diag(\bf{1},-\bf{1})$ in the retarded-advanced representation and $N$ is the total number of sites.

Applying the general scheme of the virial expansion outlined in the previous section, we obtain the following expansion for $I_q$:
\be\label{Iq_expan}
I_q=I_q^{(0)}+I_q^{(1)}+\dots
\en
The first term in this series corresponds to the first term of Eq.(\ref{Z_expan}) and is given by
\be 
I_q^{(0)}=\frac{1}{N}\sum_n\lim_{\eta\to 0}\frac{\eta^{q-1}}{4}\int DQ\Str^q\left(Q_n k \Lambda\right)e^{-\eta\sum_{m}\Str\Lambda Q_m}.
\en
One can notice that the integrand can be factorized and the integrals over different supermatrices $Q_m$  can be calculated separately. Moreover all the integrals over $Q_m$ with $m\neq n$ are equal to unity due to supersymmetry. The only non-trivial integral over $Q_n$ reproduces the standard expression for the moments of eigenvectors in the Wigner-Dyson RMT \cite{FM95}. Thus we find that $I_q^{(0)}$ coincides with the standard RMT result 
\be\label{Iq0}
I_q^{(0)}=\Gamma(q+1).
\en
Physically it corresponds to the statistics of the eigenfunctions in a system of completely decoupled metallic grains.  

The second term of the expansion (\ref{Iq_expan}) corresponding to the second term in Eq.(\ref{Z_expan}) reads
\be
\hspace*{-40pt}I_q^{(1)}=\sum_n\sum_{m>p}\lim_{\eta\to 0}\frac{\eta^{q-1}}{4N}\int DQ\Str^q\left(Q_n k \Lambda\right)\left(e^{-S[Q_m,Q_p]}-1\right)e^{-\eta\sum_{m}\Str\Lambda Q_m}
\en  
Again one can notice that, if $m\neq n$ and $p\neq n$, then the double integral over $Q_m$ and $Q_p$ is determined solely by the anomalous contributions and hence equals to zero. Thus the terms with $m=n$ or $p=n$ are the only ones which should be taken into account:
\be\label{Iq1_Q}
\hspace*{-40pt} I_q^{(1)}=\sum_{m,n}\lim_{\eta\to 0}\frac{\eta^{q-1}}{4N}\int dQ_m dQ_n\Str^q\left(Q_n k \Lambda\right)\left(e^{-S[Q_m,Q_n]}-1\right)e^{-\eta\Str\Lambda (Q_m+Q_n)}
\en
Now we need to calculate the double integral in the above expression. This task is straightforward, but technically quite challenging appreciating a non-trivial parametrization of the supermatrices. Fortunately, exactly the same integral appears at an intermediate step of the derivation of the non-linear $\sigma$-model associated with the ensemble  of the banded random matrices \cite{FM_int,Haake}. It turns out that the integrals over the Grassmann variables and the phases can be calculated exactly. The next step, which simplifies evaluation of the integral significantly, is to rescale the bosonic variables $\lambda_B$ in such a way that the limit $\eta\to 0$ can be taken explicitly in the integrand. Borrowing these results we arrive at the following expression for $I_q^{(1)}$:
\begin{eqnarray}
\label{Iq1}\hspace*{-52pt}&&I_q^{(1)}=\frac{1}{N}\sum_{m,n}\left[q(q-1)\int_0^{\infty}dz\int_0^{\infty}du\: F_{mn}(z,u)-\Gamma(q+1)\right],\\
\hspace*{-52pt}&&F_{mn}(z,u)=\sqrt{\frac{\gamma}{2\pi}}z^{q-2}u^{-\frac{3}{2}}e^{-z(u+1)-\frac{\gamma}{2}\left(u+\frac{1}{u}\right)}\left[\cosh\gamma -\frac{\sinh \gamma}{2\gamma} +\left(u+\frac{1}{u}\right)\frac{\sinh \gamma}{2}\right],
\end{eqnarray}
where parameter $\gamma=4 U_{mn}/t$ is introduced in order to lighten the notation. The appearance of $\Gamma(q+1)$ in Eq.(\ref{Iq1}) is not accidental and can be traced back to the term containing $-1$ in Eq.(\ref{Iq1_Q}), which gives exactly the same contribution as the one calculated for $I_q^{(0)}$.

The expansion (\ref{Iq_expan}) makes sense only if its different terms scale as different powers of $1/t$. Therefore from the exact expression for $I_q^{(1)}$ derived above, we need to extract only the first non-trivial term in the limit $t\to \infty$ or $\gamma\to 0$. At this point our calculations depart from the $\sigma$-model derivation, which assumes the opposite limit $\gamma\to\infty$.  

It is obvious from the form of the function $F_{mn}$ that the limit $\gamma\to 0$ can not be taken before at least one of the integrations in Eq.(\ref{Iq1}) is performed. One way to deal with this problem is to evaluate the integral over $u$ exactly:
\be
\hspace*{-60pt}\int_0^{\infty}du\: F_{mn}(z,u)= e^{-z-\sqrt{\gamma(2z+\gamma)}}z^{q-2}\frac{\gamma(2z+\gamma)\cosh\gamma+(z+\gamma)\sqrt{\gamma(2z+\gamma)}\sinh\gamma}{\gamma(2z+\gamma)}.
\en 
The expansion of the last expression into the power-series in $\sqrt{\gamma}$ yields
\be\label{F_expan}
\int_0^{\infty}du\: F_{mn}(z,u)=e^{-z}z^{q-2}\left(1-\sqrt{\frac{\gamma z}{2}}+O(\gamma)\right).
\en
Substituting this result back into Eq.(\ref{Iq1}) and integrating over $z$ we obtain 
\be
I_q^{(1)}=-\frac{1}{N}\sum_{m,n}\sqrt{\frac{\gamma}{2}}q(q-1)\Gamma(q-1/2)+O(\gamma),\quad q>\frac{1}{2}.
\en
The condition $q>1/2$ is necessary for convergence of the integral at $z\to 0$.
Restoring the dependence of $\gamma$ on $U_{mn}$ and taking into account the expression for $I_q^{(0)}$ we find the final result for the moments of the eigenfunctions
\be\label{Iq_final} 
I_q=\Gamma(q+1)\left(1-\frac{\Gamma(q-1/2)}{\Gamma(q-1)}\frac{1}{N}\sum_{mn}\sqrt{\frac{2 U_{mn}}{t}}+O(t^{-1})\right),\quad q>\frac{1}{2},
\en
which is valid for an arbitrary coupling matrix $U_{mn}$.  

The appearance of the factor $\Gamma(q+1)$ corresponding to the standard RMT result suggests that the eigenfunctions can be written as a product of two contributions. One of them describes the fluctuations of the eigenfunctions inside a grain and it is given by the Wigner-Dyson RMT. Another one, represented by the series in Eq.(\ref{Iq_final})  describes a change of the  amplitudes of the wavefunctions at different grains. The same factorization occurs for the statistics of the eigenfunctions of one-dimensional and quasi-one-dimensional disordered wires and for arbitrary dimensionality in the metallic regime \cite{M00}. 

It follows from Eqs.(\ref{Iq0},\ref{Iq1},\ref{F_expan}) that the distribution function of the eigenfunction intensities is given by
\be
P(z)=\frac{d^2}{dz^2}e^{-z}\left(1-\sqrt{z}\frac{1}{N}\sum_{mn}\sqrt{\frac{2 U_{mn}}{t}}+O(t^{-1})\right).
\en
We would like to notice that the first two terms of the series presented in Eq.(\ref{Iq_final}) are in one-to-one correspondence with the expansion for the moments of the eigenfunctions  found for almost diagonal random matrices characterized by an arbitrary variance matrix $\langle |H_{mn}|^2|\rangle$ \cite{FOR09}. It implies, in particular, the same expressions for the fractal dimensions in critical ensembles, as we discuss in Section \ref{sec_power}. 
     

\section{Two-level correlation function}\label{sec_correl}

In this section we focus on the correlation function $R_2(\omega)$ measuring the correlations between two levels separated by an energy interval $\omega$
\be
R_2(\omega)=\frac{\langle\nu(E-\omega/2)\:\nu(E+\omega/2)\rangle}{\langle\nu(E)\rangle^2},
\en
where $\nu(E)$ is the density of states and the averaging over a random matrix ensemble or different disorder realizations is assumed.

It is well known that $R_2(\omega)$ can be expressed as the following correlator of $Q$ matrices \cite{Efetov, M00}
\be
\hspace*{-60pt}R_2(\omega)=\left(\frac{1}{4N}\right)^2\re  \int DQ \left[\sum_n \Str\left(Q_n k \Lambda\right)\right]^2e^{-\frac{1}{t}\sum_{m\neq p}U_{mp}\Str Q_mQ_p+\frac{i\pi\nu \omega}{2}\sum_{m}\Str\Lambda Q_m}, 
\en
where it is assumed that $\omega$ has an infinitesimal imaginary part to ensure the convergence of the integral.
This formula allows us to find the general form of the first two terms of the virial expansion (\ref{Z_expan}): 
\begin{eqnarray}
&&\hspace*{-50pt}R_2^{(0)}(\omega)=\left(\frac{1}{4N}\right)^2\re  \int DQ \left[\sum_n \Str\left(Q_n k \Lambda\right)\right]^2e^{\frac{i\pi\nu \omega}{2}\sum_{m}\Str\Lambda Q_m},\label{diag}\\
&&\hspace*{-50pt}R_2^{(1)}(\omega)=\left(\frac{1}{4N}\right)^2\re  \int DQ \left[\sum_n \Str\left(Q_n k \Lambda\right)\right]^2\sum_{m>k}F_{mk}e^{\frac{i\pi\nu \omega}{2}\sum_{m}\Str\Lambda Q_m}.\label{non-diag}
\end{eqnarray}
In the following two subsections we evaluate $R_2^{(0)}(\omega)$ and $R_2^{(1)}(\omega)$ respectively.

\subsection{Two-level correlation function in the zeroth order approximation}
Let us first calculate $R_2^{(0)}(\omega)$. To this end we separate the diagonal and the non-diagonal contributions in Eq.(\ref{diag})
\be\label{sep}
\left[\sum_n \Str\left(Q_n k \Lambda\right)\right]^2=\sum_n \Str^2\left(Q_n k \Lambda\right)+\sum_{n\neq p} \Str\left(Q_n k \Lambda\right)\Str\left(Q_p k \Lambda\right)
\en
For the diagonal contribution the integration over any $Q_m$ with $m\neq n$ gives unity due to supersymmetry. Thus we obtain $N$ equal terms which we denote by $I^{(0)}_{d}$. For the same reason, the integration over $Q_m$ with $m\neq n$ or $m\neq p$ is trivial for the non-diagonal contribution. In this case we have $N^2-N$ equal terms $I^{(0)}_{nd}$ and hence $R_2^{(0)}(\omega)$ can be written as
\begin{eqnarray}
&&R_2^{(0)}(\omega)=\left(\frac{1}{4N}\right)^2\re \left[N I^{(0)}_{d}+(N^2-N)I^{(0)}_{nd},\right]\label{R_0}\\
&&I^{(0)}_{d}=\int dQ  \Str^2\left(Q k \Lambda\right)e^{\frac{i\pi\nu \omega}{2}\Str\Lambda Q},\\
&&I^{(0)}_{nd}=\left(\int dQ \Str\left(Q k \Lambda\right)e^{\frac{i\pi\nu \omega}{2}\Str\Lambda Q}\right)^2.\label{I_0_nd}
\end{eqnarray} 
One can notice that the integral $I^{(0)}_{d}$ is exactly the same as the one, which determines $R_2(\omega)$ in the Wigner-Dyson RMT. It can be easily calculated using the parametrization of $Q$ given in \ref{app1}:
\be\label{wd}
I^{(0)}_{d}=16+\frac{i}{\Omega^2}e^{4i\Omega}\sin 4\Omega,\quad \Omega\equiv \frac{\pi\nu\omega}{4}
\en
This result corresponds to the Wigner-Dyson expression for $R_2(\omega)$  \cite{Efetov}, which is not surprising, as we assume that the statistical properties of a single grain are described by the standard RMT.

The calculation of the non-diagonal contribution  is even simpler, as according to Eq.(\ref{str_lq}) the expression for  $\Str\left(Q k \Lambda\right)$ does not contain the full set of Grassmann variables and hence the integral in Eq.(\ref{I_0_nd}) is anomalous and equals to a constant:
\be\label{I_0_nd_ans}
I^{(0)}_{nd}=16.
\en
Eq.(\ref{R_0}) shows that for $\Omega\neq 0$ it is precisely this contribution that is dominant in the thermodynamic limit $N\to \infty$. Additionally we should take into account the self-correlation contribution contained in Eq.(\ref{wd}), as it survives in the thermodynamic limit, provided that $\omega N$ is kept constant. Thus we find that
\be\label{R_0_ans}
R_2^{(0)}(\omega)=1+\frac{1}{\nu N}\delta(\omega),\quad \mbox{as }\; N\to \infty. 
\en 
There is a simple way to understand the above result without doing any calculations. Indeed, although the level correlations in each grain are non-trivial and described by the Wigner-Dyson formula, the levels in different grains are assumed to be completely uncorrelated in the zeroth order of the virial expansion. For this reason, when the number of grains $N\to \infty$, all possible correlations except for the self-correlation vanish and $\langle\nu(E-\omega/2)\:\nu(E+\omega/2)\rangle=\langle\nu(E-\omega/2)\rangle\langle\nu(E+\omega/2)\rangle$, which leads to the trivial result (\ref{R_0_ans}), provided that the density of states in an energy interval $[E-\omega/2,E+\omega/2]$ can be approximated by a constant.
 
\subsection{Two-level correlation function in the first order approximation}

The first step in calculation of  $R_2^{(1)}(\omega)$ is to use Eq.(\ref{sep}) to separate the diagonal and the non-diagonal contributions in Eq.(\ref{non-diag}). Then one can realize that the diagonal terms give a non-zero result only, if $n=m$ or $n=k$. All other terms are anomalous and vanish, as $F_{mk}=0$ at $Q_m=Q_k=\Lambda$. In a similar way, for the non-diagonal contribution we should keep only the terms with $n=m$ and $p=k$ or $n=k$ and $p=m$. Thus we need to deal with integrals of two types:
\begin{eqnarray}
&&R_2^{(1)}(\omega)=\left(\frac{1}{4N}\right)^2 2\re \sum_{m>k}\left[ I^{(1)}_{d}+I^{(1)}_{nd}\right],\label{R_1}\\
&&I^{(1)}_{d}=\int dQ dQ^{\prime} \Str^2\left(Q k \Lambda\right)\left(e^{-b\Str Q\pQ}-1\right) e^{\frac{i\pi\nu \omega}{2}\Str\Lambda (Q+Q^{\prime})},\label{I_1_d}\\
&&I^{(1)}_{nd}=\int dQ dQ^{\prime} \Str\left(Q k \Lambda\right)\Str\left(\pQ k \Lambda\right)\left(e^{-b\Str Q\pQ}-1\right) e^{\frac{i\pi\nu \omega}{2}\Str\Lambda (Q+Q^{\prime})}\label{I_1_nd},
\end{eqnarray} 
where the dependence of $I^{(1)}_{d}$ and $I^{(1)}_{nd}$ on $m$ and $k$ is hidden in the parameter $b=2 U_{mk}/t$. Calculation of the above integrals is more challenging problem than the one we were faced in the previous section, as now we are not allowed to consider the limit $\omega\to 0$. The main steps of the calculations are outlined below and the details are presented in \ref{app2}.

We start with integration over the Grassmann variables and the phases in a similar way as it was done in the previous section. Then we are left with a sum of four-fold integrals over bosonic and fermionic variables $\lb$, $\lbp$ and $\lf$, $\lfp$. The only term which couples the bosonic with the fermionic variables originates from the measure of integration. It can be represented as an integral over an auxiliary variable of the exponential function. In this way the  bosonic and the fermionic integrals can be completely decoupled and considered separately.

The next step is to recall that we are interested in the strong coupling limit $t\to \infty$ or $b\to 0$. One can show that in this limit $R_2^{(1)}(\omega)$ is given by the asymptotic series
\be\label{series}
R_2^{(1)}(\omega)=\sum_{k\ge 0}\left(\sqrt{b}\right)^k f_k\left(\frac{\omega}{\sqrt{b}}\right).
\en 
The higher order terms of the virial expansion $R_2^{(i)}(\omega)$ have similar representations, but with $k\ge i-1$. Thus only the first term of the series (\ref{series}) is relevant in the first order approximation. For this reason, we are allowed to take the limit $b\to 0$ keeping $\omega/\sqrt{b}$ constant.

It turns out that the fermionic integrals can be easily calculated in this limit by expanding the corresponding integrands into the power series. All of them give just trivial constant contributions. The bosonic integrals are more tricky due to the fact that the bosonic variables are non-compact. Rescaling the integration variables and keeping only the leading terms of the expansion (\ref{series}) we arrive at two-fold integrals which can be evaluated exactly and gives the following expression for $R_2^{(1)}(\omega)$:
\be\label{R_1_ans}
R_2^{(1)}(\omega)=-\frac{2}{N^2}\sum_{m>k}\left[1-\frac{\pi^{\frac{3}{2}}\nu\omega\sqrt{t}}{\sqrt{8U_{mk}}}\:
e^{-\frac{(\pi\nu \omega)^2t}{8U_{mk}}}\:\erfi \left(\frac{\pi\nu \omega\sqrt{t}}{\sqrt{8U_{mk}}}\right) \right].
\en
Adding this contribution to the formula for $R_2^{(0)}(\omega)$ we obtain the final result for $R_2(\omega)$
\be
R_2(\omega)=\frac{1}{\nu N}\delta(\omega)+\frac{2}{N^2}\sum_{m>k}\frac{\pi^{\frac{3}{2}}\nu\omega\sqrt{t}}{\sqrt{8U_{mk}}}\:
e^{-\frac{(\pi\nu \omega)^2t}{8U_{mk}}}\:\erfi \left(\frac{\pi\nu \omega\sqrt{t}}{\sqrt{8U_{mk}}}\right).
\en
This is the main result of this section. Similarly to Eq.(\ref{Iq_final}) it can be used to compute $R_2(\omega)$ for the $\sigma$-model with an arbitrary coupling matrix $U$. Remarkably, the obtained expression coincides exactly with the corresponding result derived for the almost diagonal random matrices \cite{YO07}, if one identifies the coupling matrix $U_{mk}/t$ with a variance matrix $\pi^2\nu^2\langle |H_{mk}|^2|\rangle/2\langle |H_{kk}|^2 \rangle$. 

\section{Critical power-law $\sigma$-model}\label{sec_power}
In this section we apply the general expressions for $I_q$ and $R_2(\omega)$ derived in the previous sections to the $\sigma$-model showing critical behavior. One of the most studied example of this kind is the model characterized by the power-law decay of the coupling matrix elements $U_{mk}\propto |m-k|^{-\alpha}$. Such a model was derived in Ref.\cite{MFD96}
as an effective field theory corresponding to the ensemble of the power-law banded random matrices. It is believed that it  exhibits critical properties at all values of the coupling constant provided that $\alpha =2$. The mapping between the ensemble of the random matrices and the $\sigma$-model can be justified only in the large bandwidth (small coupling) limit. For this reason the critical $\sigma$-model was studied in detail in this limit \cite{MFD96, ME00, ROF11}. At the same time the model is well defined at arbitrary values of the coupling constant, in particular in the opposite regime of the strong coupling, where it can be explored by the method developed in this work. On the other hand, many analytical results are available for the original random matrix model in the case of the small bandwidth limit \cite{ME00, YK03, KOY11, MG10, BG11, EM08}. This enables us to compare the predictions obtained for the microscopic model and for the $\sigma$-model in the situation, when it seems to be unfeasible to establish a direct relation between them. 

Specifically, we consider the $\sigma$-model (\ref{partition}) with the coupling matrix
\be\label{period_coupl}
\frac{8}{t}U_{mk}=B^2\frac{\pi^2}{N^2 \sin^2\left(\frac{\pi (m-k)}{N}\right)}.
\en
The periodic form of the coupling matrix (\ref{period_coupl}) is more convenient for analytical calculations.

In order to compute $I_q$ we need to plug the coupling matrix (\ref{period_coupl}) into the general expression (\ref{Iq_final}). Calculating the sum over $m$ and $n$ in the leading order in $N$ we obtain the following expression
\be
I_q=\Gamma(q+1)\left(1-B\frac{\Gamma(q-1/2)}{\Gamma(q-1)}\ln N+\dots\right),\quad q>\frac{1}{2}.
\en
Comparing it with the equation $I_q\propto N^{-d_q(q-1)}$ defining the fractal dimensions $d_q$ we find the formula for $d_q$ in the lowest order in $B$: 
\be
d_q=B \frac{\Gamma(q-1/2)}{\Gamma(q)},
\en
which is the same as the result that one would obtain for the power-law random model with $\langle |H_{mm}|^2\rangle=1$ and 
\be
\langle |H_{mk}|^2\rangle=\left(\frac{B}{\sqrt{2\pi}}\right)^2\frac{\pi^2}{N^2 \sin^2\left(\frac{\pi (m-k)}{N}\right)}, \quad m\neq k .
\en
We conclude that identifying the coupling matrix $U_{mk}/t$ in the $\sigma$-model with the expression $\pi \langle |H_{mk}|^2\rangle/4$ in the random matrix model, we obtain a one-to-one correspondence between two results.

Next we calculate $R_2(\omega)$ for the same model. To this end we need to evaluate the double sum in Eq.(\ref{R_1_ans}) for $U_{mk}$ given by Eq.(\ref{period_coupl}) in the limit $N\to \infty$. In this limit one can first introduce rescaled variables $s=N\nu\omega$, $x=\sin(\pi|m-k|/N)$, $\phi=\pi m/N$ and then replace the sums by the corresponding integrals:
\be 
R_2^{(1)}(s)=-1+\frac{2s}{B\pi^{\frac{3}{2}}}\int_0^1 d\phi\int_0^{\sin\phi}dx \frac{x}{\sqrt{1-x^2}}e^{-\frac{s^2x^2}{b^2}}\erfi\left(\frac{sx}{b}\right).
\en
Integrating by parts over $\phi$ we are left with an integral over $x$, which can be evaluated exactly:
\be
R_2^{(1)}(s)=-e^{-\frac{s^2}{B^2}}.
\en
Finally, taking into account the contribution of $R_2^{(0)}(s)$ we obtain
\be
R_2(s)=\delta(s)+1-e^{-\frac{s^2}{B^2}}.
\en
Again we notice that this expression is in one-to-one correspondence with the random matrix result, provided that we use the same identification rule $U_{mk}/t\to \pi \langle |H_{mk}|^2\rangle/4$ as before.

\section{Conclusions}\label{conclusions}

The explicit calculations performed in the previous sections show that the first order results both for $I_q$ and $R_2(\omega)$ are equivalent to the corresponding expressions derived for random matrix models using the virial expansion method. The equivalence is very general and holds true for an arbitrary coupling matrix $U_{mk}$. In particular, we demonstrated that there is one-to-one correspondence between these two quantities calculated in the ensemble of the power-law banded random matrices at $b\ll 1$ and their counterparts calculated in the $\sigma$-model. Such a correspondence is rather unexpected taking into account the fact, that the standard mapping of the the power-law banded random matrix ensemble onto the $\sigma$-model can be justified only in the opposite limit $b\gg 1$.
 
A natural question, which appears in this context, is whether the observed equivalence can be extended to higher order terms and if it is the case, then what is the reason behind it? Below we show that under rather general assumptions both the $\sigma$-models and the random matrix models can be mapped onto the same effective field theory, so that such an equivalence is preserved at any order of the virial expansion. 

In Ref.\cite{OK06} it was noticed, that one can formulate a supersymmetric field-theoretical description of disordered quantum systems or random matrix models, which is similar to the non-linear $\sigma$-model, but it is exact and free from the saddle-point approximation. The form of the action of the new theory is similar to the action of the $\sigma$-model (\ref{partition}), but the target space of the new supermatrix $\hq$ is different from the target space of $Q$. This approach was subsequently applied to the almost diagonal random matrices in Ref.\cite{YO07}. In particular, it was shown in \cite{YO07} that in the strong coupling limit the target space of $\hq$ can be reduced by the constraint $\Str\:\hq=0$, provided that the large scale approximation can be used. The latter assumes that an exact form of the coupling matrix $U_{km}$ at small distances $|k-m|$ plays no role, and this condition is automatically satisfied for a large class of the coupling matrices in the thermodynamic limit $N\to \infty$.

The constraint  $\Str\:\hq=0$ implies that only two real and four Grassmann variables are required for parametrization of $\hq$. Its block structure in the retarded-advanced representation is similar to the block structure of $Q$ given by Eq.(\ref{q-param}):
\be\label{hq-param}
\hq=\left(\begin{array}{cc} U&0\\0&V\end{array}\right)\left(\begin{array}{cc} \td_{RR}&\td_{RA}\\\td_{AR}&\td_{AA}\end{array}\right)\left(\begin{array}{cc} U^{-1}&0\\0&V^{-1}\end{array}\right),
\en
where $U$ and $V$ are exactly the same matrices as those that involved in the Efetov's parametrization of $Q$ and defined by Eq.(\ref{UV})\cite{OK06}. Each of the matrix  $\td_{\alpha\beta}$ has only one non-zero matrix element:
\begin{eqnarray}\label{td_blocks}
\td_{RR}=\diag(\tl,0),\quad \td_{RA}=\diag(-\tl e^{-i\phi},0),\nonumber\\
\td_{AR}=\diag(\tl e^{i\phi},0),\quad \td_{RR}=\diag(-\tl,0),\nonumber\\
\tl\in[0,\infty),\quad \phi\in[0,2\pi).
\end{eqnarray} 
Now let us return to the virial expansion of the $\sigma$-model. To be specific we focus on $R_2(\omega)$, but our consideration can be applied for any other correlation function. The $n$th term of the virial expansion involves integration over $n$ supermatrices $Q_k$ and it can be written as the following asymptotic series:
\be
R_2^{(n)}(\omega)=\sum_{k\ge 0}\left(\sqrt{t}\right)^{-(k+n-1)} f_k^{(n)}\left(\sqrt{t}\omega\right).
\en
Using the same arguments as those, which were employed in the discussion of the validity of the approximation   $\Str\:\hq=0$ (see Ref.\cite{YO07} for details), one can show that actually all the terms with $k\ge 1$ can be neglected, provided that the large scale approximation is valid. In \ref{app3} we discuss the validity of this approximation for the power-law $\sigma$-model. 

In order to find the first term of the series, we scale the variables $\tl=\lambda_B/\sqrt{t}$ and $\tilde{\omega}=\sqrt{t}\omega$ and take the limit $t\to \infty$. In this limit the integrand preserves its form, but the matrices $Q_k$ should be replaced by
\be
\tq_k=\lim_{t\to\infty}\frac{1}{\sqrt{t}}Q_k\left.\right|_{\tl\; \mathrm{is\; constant}}.
\en
Looking at the parametrization of $Q$ (\ref{q-param}-\ref{D_blocks}) we can immediately notice that $\tq$ can be obtained from $Q$ by setting the variables $\lf$ and $\mf$ equal to zero and by replacing $\mb$ by $\lb$. Identifying $\phi_B$ with $\phi$ from Eq.(\ref{td_blocks}) we conclude that $\tq=\hq$. The integrands for both models are the same and the measure can be written as
\be 
d\hq=d\tq=2\frac{d\tl}{\tl^2}\frac{d\phi}{2\pi} d\rho\: d\rho^{\ast}d i\sigma\: d i\sigma^{\ast}.
\en
Since the above argument holds true at any order of the virial expansion we arrive at the conclusion that in the strong coupling limit and under the assumption of the large scale approximation both models can be mapped onto the same supersymmetric field theory. The action of this theory is the same as the one of the $\sigma$-model and the field variables $\hq_k$ are parametrized according to Eqs.(\ref{hq-param},\ref{td_blocks}). In particular, it implies that all the results derived for the almost diagonal random matrix models are applicable for the corresponding $\sigma$-models in the strong coupling limit and vice versa.

\ack
I have benefitted from discussions with Yan Fyodorov, Vladimir Kravtsov and Oleg Yevtushenko. 
I acknowledge support from the Engineering and Physical Sciences Research Council, grant 
number EP/G055769/1 and hospitality of the Abdus Salam ICTP.

\appendix 
\section{Parametrization of supermatrix Q}\label{app1}

We employ the standard parametrization of $Q$ introduced by Efetov \cite{E83} and adopt the notation used in \cite{Haake}. In the unitary symmetry class ($\beta=2$) $Q$ has the following block structure in the retarded-advanced representation:
\be\label{q-param}
Q=\left(\begin{array}{cc} U&0\\0&V\end{array}\right)\left(\begin{array}{cc} D_{RR}&D_{RA}\\D_{AR}&D_{AA}\end{array}\right)\left(\begin{array}{cc} U^{-1}&0\\0&V^{-1}\end{array}\right).
\en
The matrices $U$ and $V$ contain only Grassmann variables $\rho$, $\rho^{\ast}$ and $\sigma$, $\sigma^{\ast}$:
\be\label{UV}
U=\left(\begin{array}{cc} 1+\rho\rho^{\ast}/2 &\rho\\ \rho^{\ast}& 1+\rho^{\ast}\rho/2\end{array}\right)\quad 
V=\left(\begin{array}{cc} 1-\sigma\sigma^{\ast}/2 &i\sigma\\ i\sigma^{\ast}& 1-\sigma^{\ast}\sigma/2\end{array}\right).
\en
The matrices $D_{\alpha\beta}$ contain other four real variables $\lambda_B$, $\lambda_F$ and $\phi_B$, $\phi_F$:
\begin{eqnarray}\label{D_blocks}
D_{RR}=-D_{AA}=\left(\begin{array}{cc} \lb & 0\\ 0 & \lf\end{array}\right),\;\; \lb\in [1,\infty),\: \lb\in[-1,1]\nonumber\\
D_{RA}=\left(\begin{array}{cc} -\mb e^{-i\phi_B} & 0\\ 0 & \mf e^{-i\phi_F}\end{array}\right),\:  D_{AR}=\left(\begin{array}{cc} \mb e^{i\phi_B} & 0\\ 0 & \mf e^{i\phi_F}\end{array}\right),\nonumber\\
\phi_B,\:\phi_F\in [0,2\pi),\quad \mb=\sqrt{\lb^2-1},\:\mf=\sqrt{1-\lf^2}.
\end{eqnarray}
The measure in this parametrization is given by
\be
dQ=\frac{d\lb\: d\lf}{(\lb-\lf)^2}\frac{d\phi_B\: d\phi_F}{(2\pi)^2} d\rho\: d\rho^{\ast}d i\sigma\: d i\sigma^{\ast}.
\en
Using the expressions for the matrices $k=\diag(1,-1,1,-1)$ and $\Lambda=\diag(1,1,-1,-1)$ and the above representation of $Q$, one can easily calculate $\Str(\Lambda Q)$ and $\Str(k\Lambda Q)$:
\begin{eqnarray}\label{str_lq}
\Str(\Lambda Q)&=&2(\lb-\lf),\nonumber\\
 \Str(k\Lambda Q)&=&2\left[\lb+\lf+(\lb-\lf)(\rho\rho^{\ast}-\sigma\sigma^{\ast})\right].
\end{eqnarray}
\section{Two-level correlation function in the first order approximation}\label{app2}

According to Eq.(\ref{I_1_d}) the diagonal contribution in the first order of the virial expansion can be written as
\begin{eqnarray}
&&I^{(1)}_{d}=T(b)-T(0),\label{T_diff}\\
&&T(b)=\int dQ  \Str^2\left(Q k \Lambda\right) e^{2 i\Omega\Str\Lambda Q}Y(Q),\quad \Omega\equiv \frac{\pi\nu\omega}{4},\\
&& Y(Q)=\int d\pQ e^{-b\Str Q\pQ+2 i\Omega\Str\Lambda \pQ}.\label{Y}
\end{eqnarray}
Changing the integration variable $\pQ$ by $U^{-1}\pQ U$ in the integral defining the function $Y(Q)$, one can show that $Y(Q)=Y(\lb,\lf)$ \cite{FM_int, Haake}. Using this fact and the expressions for $\Str \Lambda Q$ and  $\Str Q k \Lambda$ from \ref{app1}, we obtain
\begin{eqnarray}
&&\hspace*{-60pt}T=4\int_1^{\infty}d\lb\int_{-1}^1d\lf\frac{1}{(\lb-\lf)^2}\int_0^{2\pi}\frac{d\phi_B d\phi_F}{(2\pi)^2}
\int d\rho^{\ast}d\rho di\sigma^{\ast}di\sigma \:e^{4i\Omega(\lb-\lf)}\times\nonumber\\
&&\hspace*{-50pt}\left[(\lb+\lf)^2+2(\lb^2-\lf^2)(\rho\rho^{\ast}-\sigma\sigma^{\ast})-2(\lb-\lf)^2\rho\rho^{\ast}\sigma\sigma^{\ast}\right]Y(\lb,\lf).
\end{eqnarray}
The three terms in the brackets in the above expression generate three contributions to $T$. The first one $T_1$ corresponding to $(\lb+\lf)^2$ is anomalous and therefore is given by
\be
T_1=16Y(1,1).
\en
The second one corresponding to $2(\lb^2-\lf^2)(\rho\rho^{\ast}-\sigma\sigma^{\ast})$ is zero, as the bosonic part of the integral converges and the fermionic one is zero. Finally, the third one $T_2$ generated by the last term in the brackets has the form
\be\label{T2}
T_2=8\int_1^{\infty}d\lb\int_{-1}^1d\lf\:e^{4i\Omega(\lb-\lf)}Y(\lb,\lf).
\en
Now we need to use the explicit form of the function $Y$, which can be found by performing integration in Eq.(\ref{Y}). The result is available in Refs.\cite{FM_int, Haake}:
\begin{eqnarray}
&&\hspace*{-60pt}Y(\lb,\lf)=e^{-2b(\lb-\lf)}+\nonumber\\
&&\hspace*{-20pt}\int_1^{\infty}d\lbp\int_{-1}^1d\lfp\:\frac{\lb-\lf}{\lbp-\lfp}e^{4i\Omega(\lbp-\lfp)-2b(\lb\lbp-\lf\lfp)}
L(\lbp,\lfp,\lb,\lf),\label{Y_ex}\\
&&\hspace*{-60pt}L(\lbp,\lfp,\lb,\lf)=2b^2\left((\lbp\lb+\lfp\lf)I_0(2b\mbp\mb)I_0(2b\mfp\mf))-\right.\nonumber\\
&&\hspace*{-20pt}\left.\left[\mb\mbp I_1(2b\mbp\mb)I_0(2b\mfp\mf))-\mf\mfp I_1(2b\mfp\mf)I_0(2b\mbp\mb))\right]\right),
\end{eqnarray}
where $I_0$ and $I_1$ are the Bessel functions. One can see that the above integral is zero at $\lb=\lf=1$ and hence $Y(1,1)=1$. Thus we find the answer for $T_1$:
\be\label{T1}
T_1=16.
\en
Another observation which can be made here is that the exponential function and the integral in Eq.(\ref{Y_ex}) generate two different contributions in Eq.(\ref{T2}), which we denote by $T_{2a}$ and $T_{2b}$ respectively. The first one can be easily calculated:
\begin{eqnarray}
&&\hspace*{-40pt}T_{2a}=8\int_1^{\infty}d\lb\int_{-1}^1d\lf\:e^{4i\Omega(\lb-\lf)}e^{-2b(\lb-\lf)}=f(\Omega,b),\nonumber\\
&&\hspace*{-40pt}f(\Omega,b)=\frac{4i\sin(4\Omega+2bi)}{(2\Omega+bi)^2}e^{4i\Omega-2b}.
\end{eqnarray}
This exact result can be represented as a sum of two different contributions: the $b$ independent part and the rest, which can be expanded into the power series:
\be \label{expan}
f(\Omega,b)=f(\Omega,0)+\sum_{k\ge 0}\left(\sqrt{b}\right)^k f_k\left(\frac{\Omega}{\sqrt{b}}\right).
\en
The explicit form of the $b$ independent part is not important as it will be canceled out later according to Eq.(\ref{T_diff}). Moreover we are interested only in first term of the power series, as the higher order terms exceed the accuracy of the first order approximation of the virial expansion:
\be\label{T2a} 
T_{2a}=\frac{i\sin(4\Omega)}{\Omega^2}e^{4i\Omega}+2\frac{b}{\Omega^2}+\dots\; .
\en   
Calculation of $T_{2b}$ requires much more efforts, as  $T_{2b}$ is given by a four-fold integral. However, one can notice that all the terms in Eq.(\ref{Y_ex}) except from the first one are factorized into the fermionic and bosonic contributions. To exploit this observation we introduce the following function
\begin{eqnarray}
&&\hspace*{-60pt}F(\Omega,\Pomega)=8\int_1^{\infty}d\lb\int_{-1}^1d\lf\int_1^{\infty}d\lbp\int_{-1}^1d\lfp\:e^{4i\Omega(\lb-\lf)+4i\Pomega(\lbp-\lfp)}\times\nonumber\\
&&\hspace*{160pt}e^{-2b(\lb\lbp-\lf\lfp)}L(\lbp,\lfp,\lb,\lf),
\end{eqnarray}
which can be used to find  $T_{2b}$
\be\label{T2b}
T_{2b}=\frac{1}{4i}\int_0^{\infty}dt \frac{\partial}{\partial \Omega}F\left.\left(\Omega,\Pomega+\frac{i}{4}t\right)\right|_{\Pomega=\Omega}.
\en 
It has an advantage that now the fermionic and the bosonic integrals do not mix up:
\begin{eqnarray}
&&\hspace*{-60pt}F(\Omega,\Pomega)=16b^2\sum_{\alpha=1}^4\int_1^{\infty}d\lb\int_{-1}^1d\lf\int_1^{\infty}d\lbp\int_{-1}^1d\lfp\:e^{4i\Omega(\lb-\lf)+4i\Pomega(\lbp-\lfp)}\times\nonumber\\
&&\hspace*{140pt}e^{-2b(\lb\lbp-\lf\lfp)}f_{\alpha}(\lb,\lbp)g_{\alpha}(\lf,\lfp),\\
&&f_1=\lbp\lb I_0(2b\mbp\mb),\quad g_1=I_0(2b\mfp\mf),\nonumber\\
&&f_2=I_0(2b\mbp\mb),\quad g_2=\lfp\lf I_0(2b\mfp\mf),\nonumber\\
&&f_3=-\mbp\mb I_0(2b\mbp\mb),\quad g_3=I_0(2b\mfp\mf),\nonumber\\
&&f_4=I_0(2b\mbp\mb),\quad g_4=\mfp\mf I_1(2b\mfp\mf).
\end{eqnarray}
Thus the problem is reduced now to the calculation of the two-fold bosonic and fermionic integrals, which we denote by $A_{\alpha}$ and $B_{\alpha}$ respectively:
\be\label{F}
F(\Omega,\Pomega)=16b^2\sum_{\alpha=1}^4 A_{\alpha}B_{\alpha}.
\en 
Again we are interested only in the leading term of the expansion  (\ref{expan}) and therefore we are allowed to consider the limit $b\to 0$ keeping the ratios $\Omega/\sqrt{b}$ and $\Pomega/\sqrt{b}$ constant. In this limit the fermionic integrals can be easily evaluated by expanding the integrands into the power series and the results read 
\be\label{AB}
B_1=B_3=4+O(b),\quad B_2=B_4=O(b).
\en
The bosonic integrals are less trivial due to the non-compact character of the bosonic variables. In order to expand the integrals into the power series we introduce new variables $u=2b\lb\lbp$,  $v=\lb/\lbp$, $\eta=\Omega/\sqrt{b}$ and $\etap=\Pomega/\sqrt{b}$. Then keeping only the leading order in $b$ term in the integrands we obtain 
\begin{eqnarray}\label{A1A3}
A_1&=&\frac{1}{8b^2}\int_0^{\infty}du\int_0^{\infty}dv \:\frac{u}{v}\: 
e^{4i\left(\eta \sqrt{\frac{uv}{2}}+\etap\sqrt{\frac{u}{2v}}\right)}\:e^{-u}I_0(u).\nonumber\\
A_3&=&-\frac{1}{8b^2}\int_0^{\infty}du\int_0^{\infty}dv \:\frac{u}{v}\: 
e^{4i\left(\eta \sqrt{\frac{uv}{2}}+\etap\sqrt{\frac{u}{2v}}\right)}\:e^{-u}I_1(u).
\end{eqnarray}
In a similar way, one can show that $A_4$ converges while $A_2$ logarithmically diverges as $b\to 0$. According to Eq.(\ref{AB}) this implies that their contributions to $F(\Omega,\Pomega)$ can be neglected. Substituting the results of Eqs.(\ref{A1A3}, \ref{AB}, \ref{F}) into Eq.(\ref{T2b}) and integrating over $t$ and $v$ we find
\be\label{T2b_ans}
T_{2b}=-32\int_0^{\infty}du \:u K_2\left(-8i \Omega\sqrt{\frac{u}{2b}}\right)e^{-u}\left[I_1(u)-I_0(u)\right],
\en
where $K_2$ is the modified Bessel function. Collecting the results of Eqs.(\ref{T1},\ref{T2a},\ref{T2b}) we obtain
\begin{eqnarray}\label{T_ans}
T(b)&=&16+\frac{i\sin(4\Omega)}{\Omega^2}e^{4i\Omega}+2\frac{b}{\Omega^2}-\nonumber\\
&&32\int_0^{\infty}du \:u K_2\left(-8i \Omega\sqrt{\frac{u}{2b}}\right)e^{-u}\left[I_1(u)-I_0(u)\right].
\end{eqnarray}

In order to compute the non-diagonal contribution Eq.(\ref{I_1_nd}) exactly the same approach can be used. Skipping the details of the calculations, we present here the final result for $S(b)$, which plays the same role as $T(b)$ for the diagonal contribution:
\be\label{S_ans}
S(b)=16-32\int_0^{\infty}du \:u K_0\left(-8i \Omega\sqrt{\frac{u}{2b}}\right)e^{-u}\left[I_1(u)-I_0(u)\right].
\en 
It turns out that both integrals in Eq.(\ref{T_ans}) and Eq.(\ref{S_ans}) can be computed analytically and the real part of their sum is given by
\begin{eqnarray}
&&\hspace*{-60pt}\re\int_0^{\infty}du \:u \left[K_0\left(-8i \Omega\sqrt{\frac{u}{2b}}\right)+K_2\left(-8i \Omega\sqrt{\frac{u}{2b}}\right)\right]e^{-u}\left[I_1(u)-I_0(u)\right]=\nonumber\\
&&\frac{b}{16 \Omega^2}+\frac{1}{2}-\frac{\sqrt{\pi}\Omega}{\sqrt{b}}\:e^{-\frac{4\Omega^2}{b}}\:\erfi \left(\frac{2\Omega}{\sqrt{b}}\right),
\end{eqnarray}
where $\erfi(z)=(2/\sqrt{\pi})\int_0^ze^{x^2}dx$. Therefore the final expression for the sum of the diagonal and non-diagonal contributions reads
\be
\re(I_d^{(1)}+I_{nd}^{(1)})=-16\left[1-\frac{\sqrt{\pi}\Omega}{\sqrt{b}}\:e^{-\frac{4\Omega^2}{b}}\:\erfi \left(\frac{2\Omega}{\sqrt{b}}\right)\right].
\en
Substituting this expression into Eq.(\ref{R_1}) and recalling the definition of $b$ and $\Omega$ we arrive at Eq.(\ref{R_1_ans}).
\section{Validity of the large scale approximation for the power-law $\sigma$-model}\label{app3}
In this Appendix we show, on the example of the power-law $\sigma$-model, why the term of the order of $1/t$ in the expansion of $I_q^{(1)}$ can be neglected in comparison with the term of the same order originating from the expansion of $I_q^{(2)}$.

It follows from Eqs.(\ref{Iq1},\ref{F_expan}) that the next term in Eq.(\ref{Iq_final}) has the following structure:
\be
G(q)\frac{1}{N}\sum_{mn}\frac{U_{mn}}{t},
\en
where $G(q)$ is some function of $q$, which does not depend on $N$. Substituting the expression for $U_{mn}$ (\ref{period_coupl}) into this formula we obtain that in the thermodynamic limit $N\to \infty$ the result is $C G(q) B^2$ with some constant $C$.

On the other hand the leading order in $B$ term for  $I_q^{(2)}$ will be the same as the one calculated for the almost diagonal random matrices and it is given according to Ref.\cite{KOY11} by the following expression
\be
B^2(\tilde{G}_1(q)\ln^2 N +\tilde{G}_2(q)\ln N+\dots),
\en
with some $N$-independent functions $\tilde{G}_1(q)$ and $\tilde{G}_2(q)$. One can see that although this term has the same order in $B$ it will be dominant in the thermodynamic limit $N\to \infty$. The reason for a different $N$-dependence appearing here is related to the fact, that in  $I_q^{(2)}$ there is an additional summation over the indices of the coupling matrix 
$\sum_{mnp}\sqrt{\frac{U_{mn}}{t}}\sqrt{\frac{U_{np}}{t}}$. The importance of such an additional summation for the justification of the large scale approximation was emphasized in Ref.\cite{YO07} 

\end{document}